# Spectroscopic flat-fields can be used for precision CCD gain and noise tests

J. Gordon Robertson[1,2]*
[1]Sydney Institute for Astronomy, School of Physics, University of Sydney, NSW 2006, Australia
[2]Australian Astronomical Optics, Macquarie University, 105 Delhi Rd, North Ryde, NSW 2113, Australia

**Abstract**
One of the basic parameters of a CCD camera is its gain, *i.e.* the number of detected electrons per output Analogue to Digital Unit (ADU). This is normally determined by finding the statistical variances from a series of flat-field exposures with nearly constant levels over substantial areas, and making use of the fact that photon (Poisson) noise has variance equal to the mean. However, when a CCD has been installed in a spectroscopic instrument fed by numerous optical fibres, or with an echelle format, it is no longer possible to obtain illumination that is constant over large areas. Instead of making do with selected small areas, it is shown here that the wide variation of signal level in a spectroscopic 'flat-field' can be used to obtain accurate values of the CCD gain, needing only a matched pair of exposures (that differ in their realisation of the noise). Once the gain is known, the CCD readout noise (in electrons) is easily found from a pair of bias frames. Spatial stability of the image in the two flat-fields is important, although correction of minor shifts is shown to be possible, at the expense of further analysis.

**Keywords:** astronomical instrumentation — spectrographs — CCD gain — gain and noise – data analysis and techniques

## 1 INTRODUCTION

The gain of an installed CCD represents the number of electrons (detected photons) per analogue to digital unit (ADU) as produced by the system's signal chain and analogue to digital converter. The gain must be known in order to compare the detected photon rate to the incident flux of a standard star and hence find the throughput and sensitivity. It is also needed in order to measure the system's readout noise in electrons (and then to show that the readout noise has been adequately sampled by the digitisation). Further, ongoing monitoring of an installed system's gain serves to check for any drifts in the data-handling electronics. This is particularly important where high precision spectroscopy or spectrophotometry is to be performed.

The well-known method of finding the gain is to use the fact that photon noise obeys Poisson statistics, with variance equal to the mean. An exposure having uniform illumination over a substantial area of the detector can be used to assess the pixel-to-pixel sample variance $s^2$ in ADU$^2$; similar exposures at other illumination levels can then be used to plot $s^2$ *vs* the sample mean $\bar{x}$. (The bias level must be subtracted from the mean so that $\bar{x}$ is proportional to the number of detected photons.) The result is a straight line plot whose slope gives the reciprocal gain, and intercept gives the readout noise (*e.g.* Janesick et al. 1987; Janesick 2007):

$$s^2 = \frac{\bar{x}}{g} + \frac{\sigma_{\mathrm{ro}}^2}{g^2} \qquad (1)$$

where $g$ is the gain in e$^-$/ADU and $\sigma_{\mathrm{ro}}$ is the rms readout noise in electrons.[1]

The above procedure can be readily applied in laboratory tests or for imaging instruments with a suitable flat-field illumination. But it is difficult to apply this procedure for a CCD which has been installed in a fibre-fed spectroscopic instrument, or one with an echelle format. In this case, 'flat-field' illumination refers to the use of a continuum lamp which outlines the spectrum from each fibre (or echelle order) on the detector, but with large intensity variations across the fibres and to a lesser extent along them (*e.g.* Figure 1). Procedures have been devised to use a number of small areas of nearly constant illumination (*e.g.* Wilson et al. 2019), but the far smaller number of pixels and variation of intensity even across small areas limit the precision of that technique.

---

[1]In practice it is necessary to alleviate the effects of small illumination gradients by taking the difference of two exposures at the same level and then allowing for the doubling of variance, and also to clip off discrepant values due to cosmic rays or defective pixels.

*Gordon.Robertson@sydney.edu.au

This work takes a different approach, which enables spectroscopic flat-fields to be used in their entirety, resulting in accurate values of the detector gain.

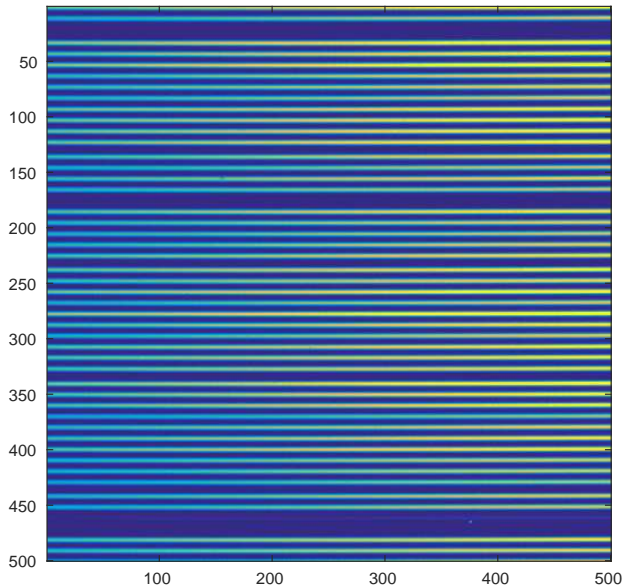

**Figure 1.** Example of a $500 \times 500$ pixel subsection from a spectroscopic flat-field of a fibre-fed instrument. (This is part of one of the flat-fields used in Section 5.2)

## 2 PROCEDURE

### 2.1 Theoretical basis

The variance of pixel values is calculated using the standard formula:

$$s^2 = \frac{1}{N-1} \sum (x_k - \bar{x})^2 \qquad (2)$$

where $N$ is the number of data points being summed over. In the standard method as described above, $N$ will be large, say $10^3 - 10^6$. The procedure introduced here uses the absolute minimum number of points, $N = 2$. The $x_k$ in this case are the values of a given pixel in two successive flat-fields. The precision of such a variance determination is low, but there are millions of pixels each providing a value of $s^2$ at their particular $\bar{x}$ (which is the mean of the values of that pixel in the two flat-field frames). The key to the success of the method is that although the mean $\bar{x}$ of two exposures for a certain pixel does not in general represent its true mean $\mu$, it is nevertheless true that for the Poisson noise component the expectation value of the associated sample variance ($s^2 g^2$ photon$^2$) is equal to the *sample* mean $\bar{x} g$, not the true mean $\mu g$. This means that the $(\bar{x}, s^2)$ pair provides an unbiased point on the plot such as Figure 2a.[2] Due to the large scatter of the individual variance values, they are grouped into bins along the $\bar{x}$ axis for plotting and parameter fitting.

In order for this procedure to give accurate results, it is important that there is no contribution to variance other than Poisson (photon) noise and readout noise. In other words the two flat-field frames used must be a matched pair. This places quite stringent limits on the stability of the spectral image between the two frames used for the test, as described in Section 5. On the other hand, this method is not affected by fixed-pattern noise (pixel-to-pixel gain variations), provided that each pixel's gain is the same in the two exposures of the pair. The method is also not troubled by scattered light between the spectral traces, as long as it remains constant − in fact it may provide useful data at the low-illumination end of the relationship. It is likely that there may be a small change in the intensity of the illumination lamp between the two exposures of the pair. This is easily overcome by finding the average intensity ratio and scaling to equality before finding the variance.

### 2.2 Analysis Steps

The steps used in the analysis here were:

1. Bias subtraction
2. Find the average ratio of the two data frames (data1/data2) and scale the input data appropriately. It is assumed that the scaling factor is close to unity (since the frames are a matched pair), hence the Poisson statistics will not be significantly disturbed.
3. Set the bin edges in $\bar{x}$, for the range over which data will be plotted.
4. For each $\bar{x}$ bin, calculate for every pixel $i$ in the bin:

$$\bar{x}_i = (x_1 + x_2)/2 \qquad (3)$$

   and specialising equation 2 to the case of $N = 2$,

$$s_i^2 = \frac{1}{2}(x_1 - x_2)^2. \qquad (4)$$

5. Reject points (pixel pairs) with excessive variance, due to bad pixels, cosmic rays etc. A simple rejection criterion is to cut off variances greater than say $(5s)^2$ where $s^2$ is the variance estimated using equation 1, with preliminary (or iterated) estimates for the gain and readout noise. A more sophisticated criterion is given in Section 4.

---

[2] For example, consider a large ensemble of pairs which have true mean = 100 photons. Taking the subset which, due to Poisson scatter, have sample mean = 95, it can be shown that the expectation value of the sample variance *of this subset* is 95, not 100.

6. Find the averages of $\bar{x}$ and $s^2$ over all the points in the particular bin $j$:

$$\bar{x}_j = \frac{1}{N_j} \sum \bar{x}_i \qquad (5)$$

and

$$s_j^2 = \frac{1}{N_j} \sum s_i^2. \qquad (6)$$

where $N_j$ is the number of pixels in bin $j$ after rejection of any outliers.

7. Find the estimated uncertainty of each binned variance (*e.g.* Barlow 1989):

$$\sigma_{s_j^2} = s_j^2 \sqrt{\frac{2}{N_j}} \qquad (7)$$

This is important because the number of pixels contributing to the bins (*i.e.* $N_j$) can vary widely.

8. Plot the binned variance *vs* mean.

9. Fit a straight line to the $\bar{x}_j, s_j^2$ binned data over the selected suitable range. It is usually necessary to omit from the fit the values of $\bar{x}$ close to zero, due to the large fractional change in $\bar{x}$ across a bin, and the effect of any small error in subtraction of the bias. The fit should allow for the differing uncertainties of the data points, and should provide the intercept and slope with uncertainties, and the reduced $\chi^2$ (*i.e.* $\chi_\nu^2$) of the fit (*e.g.* Press et al. 1992).

10. In order to see departures of the data from the straight line, plot the residuals of the data relative to the fit. This is necessary because departures may be fractionally small but statistically significant (because large $N_j$ produces small $\sigma_{s_j^2}$).

11. Other plots that may be useful include the variation of $N_j$ with $\bar{x}_j$, as well as the number of rejected points, and the cutoff variance beyond which pixels are ignored.

12. In practice, there may be some additional errors in the variances, which while fractionally small can result in a value of $\chi_\nu^2$ significantly greater than its expected value of 1, despite the line being able to give a moderately accurate value of the slope (hence the gain). In this case the uncertainty values of the slope and intercept as reported by the line fitting routine should be scaled by $\sqrt{\chi_\nu^2}$ (Press et al. 1992). If the errors lead to any form of systematic behaviour of the residuals, this scaling will still underestimate the parameter errors.

13. From the fitted intercept and slope, find the gain and readout noise from equation 1, with uncertainties from standard formulas (*e.g.* Bevington and Robinson 1992).

A Bayesian approach has not been used in this analysis for several reasons: (a) There is a great deal of data contributing to the posterior, but little useful prior information; (b) There is no benefit in obtaining the joint probability distribution showing the correlation of the gain and readout noise estimates, because the readout noise will in any case be derived separately, as described in Section 3.1; (c) With numerous bins on the $\bar{x}$ axis any improvement in precision due to an unbinned treatment would be negligible; (d) It is desirable to keep the analysis as simple as possible so it can be readily applied in practice.

# 3 APPLICATION

## 3.1 Example

Figure 2 shows an example from testing of one quadrant of the blue CCD of the GHOST high resolution echelle spectrograph, under construction for Gemini South (Sheinis et al. 2016). The data are binned between 0 and 1000 ADU, and the variance vs mean line in panel (a) fitted between 40 and 1000 ADU. This shows the straight line due to increasing Poisson noise, and intercept due to readout noise. The intercept is close to zero due to the very low readout noise of modern CCDs.

The residuals show no significant systematic errors, and the fit has $\chi_\nu^2 = 1.27$. The fitted line gives gain = $0.5835 \pm 0.0012$ e$^-$/ADU (*i.e.* 0.2% precision) and readout noise = $2.93 \pm 0.03$ e$^-$. However, the readout noise should be obtained by using the above gain figure in conjunction with the standard deviation computed from two bias frames:

$$\sigma_{\text{ro}} = \frac{g}{\sqrt{2}} \ \sigma_{\text{x}_1 - \text{x}_2} \qquad (8)$$

where $\sigma_{\text{x}_1 - \text{x}_2}$ is the standard deviation of the differences of pixels in two bias (zero exposure dark) frames over a suitable large area excluding any discrepant pixels, and the result $\sigma_{\text{ro}}$ is in e$^-$. This will give a more accurate value than the intercept of the fitted line.

## 3.2 Simulations

A number of simulations have been performed to check the validity of the above analysis method. One of the two flat-fields from the above GHOST example data was used as a template for the distribution of pixel intensities. For the purpose of the simulation, this data can be regarded as giving the exact mean values $\mu$, then from this two frames of test data were calculated using the Poisson and Normal random number generators in MATLAB[3], for the photon noise and readout noise respectively. The values of gain and readout noise used for generating the data are thus known exactly. The resulting two frames were analysed as before. The results showed that gain and readout noise are recovered within $1-2\sigma$ of the true values, and $\chi_\nu^2 \sim 1$. This shows that the gain and readout noise values are unbiased, and that the error estimates are realistic.

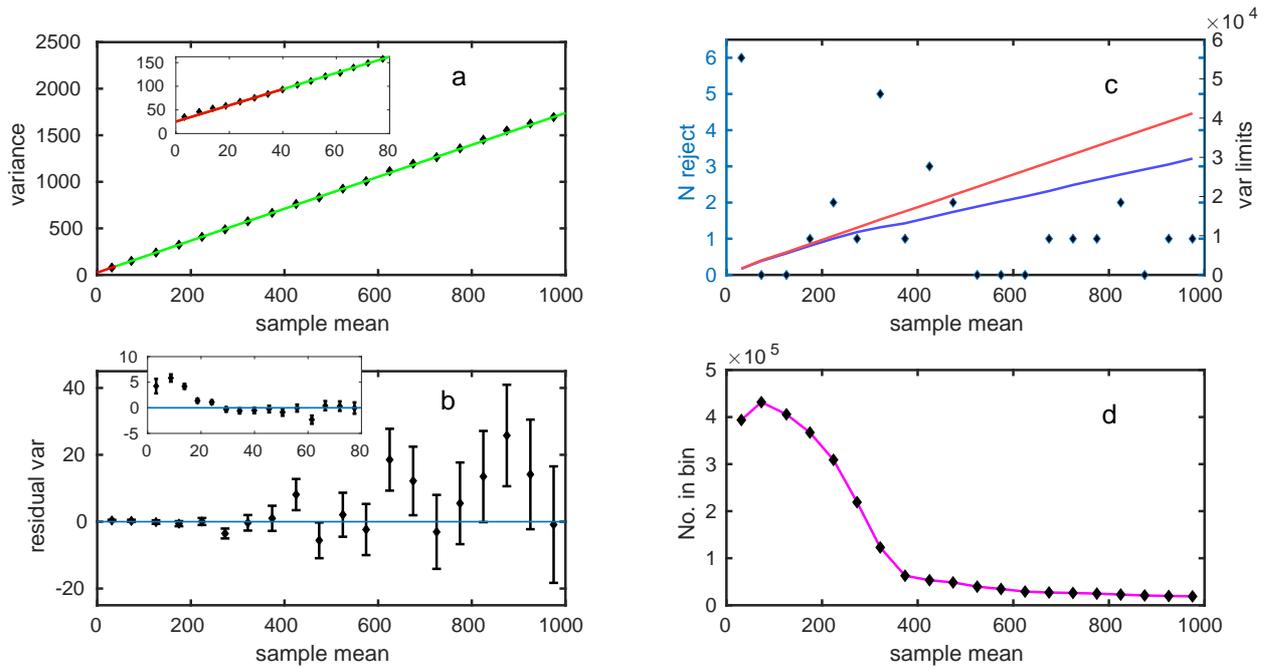

**Figure 2.** Results of analysis of the example matched pair of flat-fields, from the blue CCD of the GHOST spectrograph. The four panels show: (a) Plotted points are the binned average $\bar{x}$, $s^2$ values; the green line is the least squares fit to the points; the red segment is an extrapolation of the line beyond the points used for fitting. The inset shows the region of low sample means, with finer binning. (b) The residuals of the variance values relative to the fitted line, with $1\sigma$ error bars. The inset again shows the region of low $\bar{x}$. (c) Points show the number of pixels rejected in each bin (left hand scale); the red line shows the variance cutoff from the simple $5\sigma$ rejection criterion (right-hand scale), while the blue line shows the probability-based criterion as used in this analysis. (d) The number of pixels contributing to each bin (note the $\times 10^5$ scale multiplier). The number varies substantially with mean signal level, due to the nature of the illumination and the format of the spectral image.

## 4 VARIANCE DISTRIBUTION AND REJECTION CRITERION

While the expectation of the sample variance computed from $N = 2$ realisations at each pixel gives an unbiased point on the variance *vs* mean relation, it is nevertheless the case that the distribution of the pixel variance values $s_i^2$ (in any narrow range of $\bar{x}$ values) has a very skew form. Figure 3 shows a histogram of the values of pixel variances for the example data from Sec. 3.1. Low values are overwhelmingly more common, with the distribution dropping exponentially over most of the range, and even more rapidly at low variance. This is expected, because the variances calculated from equation 4 have close to a $\chi^2$ distribution with one degree of freedom. The theoretical form of the probability distribution function in this case is

$$f(Z) = \frac{1}{2\sqrt{\pi}} \left(\frac{Z}{2}\right)^{-\frac{1}{2}} \exp\left(-\frac{Z}{2}\right) \qquad (9)$$

(*e.g* Eadie et al. 1971; Stuart and Ord, 1987). The observed distribution of variance values will follow this form to the extent that $(x_1 - x_2)$ has a normal distribution. At large $\bar{x}$ values this will be a good approximation but deviations can be expected for lower $\bar{x}$. The exact form is not important here, since the distribution is to be used only to set a rejection criterion for outlier variances.

In order to relate this theoretical probability distribution to the data as plotted, it is necessary to scale the independent variable, such that

$$Z = gV/\bar{x} \qquad (10)$$

where $V$ is the variance value for the computation, and to normalise by multiplying the function given in equation 9 by a further factor $gN_jW_V/\bar{x}$ where $N_j$ is the total number of pixels contributing to the histogram, and $W_V$ is the width of the bins on the variance axis (assuming constant bin widths are used). A preliminary (or iterated) value for the gain $g$ is required.

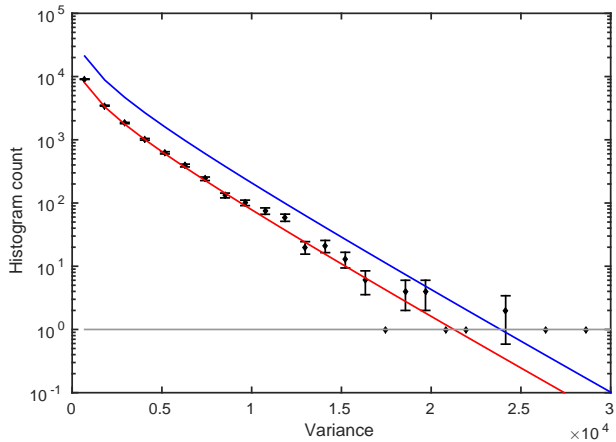

**Figure 3.** Histogram of the pixel variance values from the GHOST example data, for $\bar{x}$ values between 800 and 850, illustrating the strongly skew nature of the sampling distribution. The red curve shows the theoretical distribution as modelled by a scaled $\chi^2$ distribution with one degree of freedom; the blue curve shows the approximated integral of this theoretical curve from each value of the abscissa to infinity (*i.e.* the expected number of pixels above that value of the variance). Where the blue integrated curve crosses the horizontal grey line at a count of unity is a suitable variance value for the outlier cutoff.

The red line plotted on Figure 3 shows this theoretical distribution is a good fit to the observed histogram points.

A reasonable criterion for rejection of outlier variance values is to find the variance limit beyond which say 1 pixel pair might be expected on the basis of the theoretical model. It is immaterial to reject one valid point out of some $10^4 - 10^5$, and this criterion will catch the large outliers. Implementation requires integration of the above $\chi^2$ distribution from each value of variance to infinity, then location of the variance limit at which this number falls to unity. As Figure 3 shows, the histogram values in the vicinity of the cutoff point are falling close to exponentially, *i.e* the variation of the $(Z/2)^{-1/2}$ factor is no longer significant. For the present purpose it is therefore sufficient to treat that factor as constant (equal to its value at the lower limit of the integral) and integrate only the exponential factor. The result is

$$n_c = N_j \sqrt{\frac{2\bar{x}}{\pi g V_{\lim}}} \exp\left(-\frac{g V_{\lim}}{2\bar{x}}\right) \quad (11)$$

where $n_c$ is the estimated total number of counts above the variance $V_{\lim}$ and is shown as the blue curve in Figure 3. The rejection criterion is to omit all variances above the value $V_{\lim}$ which gives $n_c = 1$ (an approxi-mate numerical solution of the equation is adequate). This criterion has the advantage that it adapts to differing total numbers of pixels included in the $\bar{x}$ bin.

# 5 IMAGE STABILITY REQUIREMENTS

It is clear that the success of the method developed here depends on the true mean illumination for each pixel being the same for the two members of the matched pair of flat-fields, which should differ only in their realisations of the photon and readout noises. Since the illumination is very non-uniform across the CCD (being dispersed images of individual fibres or echelle orders), even a slight shift of the spectra between the two frames can give spurious contributions to the variance. This section considers the resulting stability requirements, and illustrates with an example showing that small spatial shifts can be mitigated. Variations in the intensity of the illumination lamp are not considered, because they should be removed as part of the routine analysis (Section 2.2).

## 5.1 Effect of image motion

The precise effects of a given image shift between frames 1 and 2 of the pair will clearly depend on the detailed form of the spectral images, especially the width of the spectra in the spatial direction. Shifts in the spectral direction are likely to cause much lower levels of anomalous variance than shifts in the spatial direction.

A simple model can be used to find indicative values of pixel shifts that could cause problems. Errors due to spatial shifts depend on the slope of the pixel intensities across the pixel of interest. The model considers a fibre image in the spatial direction as a peak with sides of constant slope, *i.e.* a triangle. Let its Full Width at Half Maximum (FWHM) be $\Gamma$ pixels, and its peak height $P$ ADU. Consider a spatial shift of $\delta$, measured as a fraction of a pixel. The intensity error $\Delta i$ introduced as the shift slides the pixel up or down the sloping profile is

$$\Delta i = P\delta/\Gamma. \quad (12)$$

When the pixel variance is calculated using equation 4 this will cause an additional variance of

$$V_\delta = \frac{1}{2}\Delta i^2 = \frac{1}{2}\left(\frac{P\delta}{\Gamma}\right)^2. \quad (13)$$

When averaged over all pixels across the spatial profile of the fibre track, the measured value $\bar{x}$ will be about half the peak value , *i.e.* $P \approx 2\bar{x}$.

A reasonable criterion for the spatial shift $\delta_{\lim}$ at which the effects become serious is to set the additional variance due to the spatial shift equal to the $1\sigma$ uncertainty of the Poisson variance (equation 7). This leads to

$$\delta_{\lim} = \Gamma(g\bar{x})^{-1/2}(2N_j)^{-1/4}. \quad (14)$$

As expected, the limit on shifts $\delta_{\text{lim}}$ is tighter for narrower features (smaller $\Gamma$) and for pixels recording higher intensities ($\bar{x}$). There is only a weak dependence on the number of pixels $N_j$ contributing to the particular $\bar{x}$ bin. As an example, if the fibre tracks are 3 pixels wide at FWHM, gain is $0.6 \, \text{e}^-/\text{ADU}$, $N_j = 2 \times 10^5$ and $\bar{x} = 300$, then $\delta_{\text{lim}} = 0.009$, which is a quite stringent stability requirement.

This approximate indicative calculation shows the importance of spatial stability, and that it is particularly important at higher values of pixel mean $\bar{x}$.

### 5.2 A second example

Figure 4 shows the result of analysis of two flat-field frames from the AAOmega spectrograph at the Anglo-Australian Telescope (Saunders et al. 2004). The fitted line (over the range $20 - 300$ ADU) gives gain = $1.77 \pm 0.02 \, \text{e}^-/\text{ADU}$, readout noise $2.9 \pm 0.5 \, \text{e}^-$ and $\chi^2_\nu = 49.6$. The high $\chi^2_\nu$ shows that the fit is not ideal. The calculated $\sim 1.3\%$ precision of the gain (which includes scaling by $\sqrt{\chi^2_\nu}$, Sec. 2.2) may underestimate the actual uncertainty, due to systematic errors. The larger uncertainty of the gain leads to significant uncertainty in the intercept, and the readout noise would be much better determined by using the gain value in conjunction with a pair of bias frames (equation 8).

These two flat-field frames were not acquired with the specific aim of gain and noise tests, and some spatial instability can be suspected as a possible cause of the imperfect fit. (Unlike GHOST which has a fixed format, AAOmega has adjustable grating tilt and an articulated camera.) The rise of residual variance for higher values of the sample mean is consistent with additional variance from a small spatial mismatch.

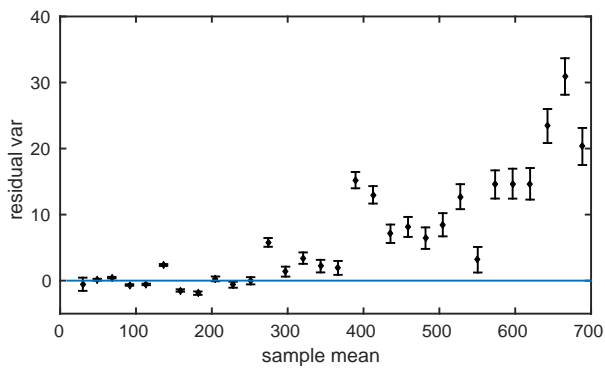

**Figure 4.** Residual variance about the fitted line for the AAOmega flat-field pair. The $s^2 \; vs \; \bar{x}$ line has been fitted over the range 20 - 300 ADU.

In order to diagnose the presence of spatial shifts between a pair of flat-fields, a useful quantity to calculate

(for every pixel) is

$$Q_i = \sqrt{g} \, \frac{(x_1 - x_2)}{\sqrt{x_1 + x_2}} \qquad (15)$$

The numerator is sensitive to shifts, and will show a characteristic +/- signature across a shifted fibre profile. The denominator provides normalisation so that Poisson noise in high signal areas is not mistaken for a shift. For Poisson noise $Q$ is distributed with a mean of zero and standard deviation of unity (except for values of $\bar{x}$ close to zero). Inspection of the $Q$ image from this pair of flat-fields did show distinct traces of the fibre spectra; Figure 5 shows this quantity averaged over 100 columns. Poisson noise would be expected to give an unstructured plot with standard deviation $\sim 0.1$ after averaging, but the Figure shows clear indications of the effect of some spatial shift.

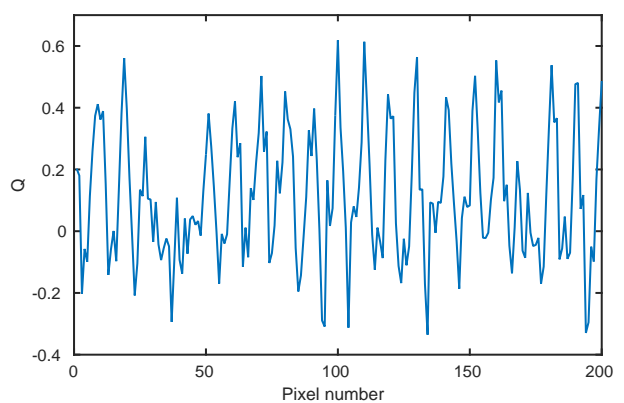

**Figure 5.** The $Q$ statistic diagnoses the presence of a small spatial shift between the two flat-fields of a pair. It has been averaged along the fibre spectra for 100 columns. The characteristic +/- signature when crossing individual fibres is seen.

### 5.3 Correction for spatial shift

The following procedure was found to be successful in characterising and then largely removing the effect of the spatial shift, *i.e.* bringing the two flat-fields into alignment. Only the shift component perpendicular to the spectral traces was examined, since shifts along the smooth flat-field spectra have little effect.

1. The flat-field frames were divided into $100 \times 100$ pixel blocks.
2. In each block the pixel intensities were averaged along the spectral rows.
3. The $Q$ parameter was calculated, giving a 1D scan similar to Figure 5.
4. Using spline interpolation, the scan of pixel intensities from the second flat-field was moved along

the direction perpendicular to the spectral traces in 11 steps over a range of $\pm 0.03$ pixels, and the standard deviation of the $Q$ scan calculated for each shift step. A 4th degree polynomial was fitted to the standard deviations and from it the pixel offset producing the minimum standard deviation of $Q$ was recorded for that position on the CCD. (It is interesting to note that the optimum offsets could not be found by cross-correlation, even with edge tapers. This is due to the dominance of edge effects, which do not affect the $Q$ parameter because it makes full use of the fact that the two scans are almost identical once correctly aligned.)

5. The array of shifts was examined and found to show a linear slope along both axes of the CCD, *i.e* the amplitudes of the shifts lay on a tilted plane. This is consistent with the shifts being due to a small rotation with a possible translation component. The 3-parameter fit to a tilted plane was found using the MATLAB function *fit*.

   The shifts were small, ranging from $-0.0017$ pixel in one corner to $+0.0186$ pixel in the opposite corner. For normal astronomical observations they would be negligible, but they are large enough to affect the variance *vs* mean method.

6. Spline interpolation was used to move each spatial column in each block of the second flat field by the amount indicated by the value of the fitted tilted plane at that position. (Note that linear interpolation cannot accurately shift a peak, since it can never give a value higher than the two values being interpolated.)

   The shifts are too small for the interpolation to cause significant noise correlations or data smoothing, so the interpretation of variances remains valid.

Figure 6 shows the results of analysis after this alignment procedure. Comparison with Figure 4 shows that the residuals have been greatly reduced, and the variance fit could now be used up to much greater pixel mean values. This fit gives gain = $1.791 \pm 0.002$ e$^-$/ADU (0.12% precision) and readout noise = $3.43 \pm 0.07$ e$^-$, with $\chi^2_\nu = 2.77$. These results improve on the predetermined values given in the FITS header, viz gain = 1.99 and readout noise = 4.25. (As noted in Section 3.1, further improvement in the readout noise precision can be obtained by using the accurate value of gain in conjunction with the variance from two bias frames.)

## 6  CONCLUSIONS

This work has shown that spectroscopic flat-fields can be used to find accurate values of the CCD gain (e$^-$/ADU), despite there being no substantial areas of the CCD with uniform intensity. The data required is a matched pair of flat-fields, *i.e.* a pair taken under identical conditions, hence differing only in the realisation of the photon and readout noises. Once the gain value has been found, the rms readout noise (in e$^-$) can be readily found from a pair of bias frames.

Although this procedure could in principle reveal any non-linearity in the CCD response, departures from the straight line variance *vs* mean relationship can be due to other causes, and the number of pixels at high mean values may in any case be inadequate for this purpose. Linearity is better checked using a stable lamp and finding the mean levels in exposures of various durations.

The principal caveat to the method presented here is the need for spatial stability, since small shifts can disturb the analysis. If flat-fields are taken specifically for the purpose of gain and noise tests, precautions should be taken to ensure temperature stability and avoidance of mechanical motion for some time preceding the tests. The two frames should be taken with minimal delay between them. Nevertheless, even with some instability a useful value of gain can still be found, and with further analysis a small spatial shift between the pair of flat-fields can be corrected.

**Acknowledgments**

I thank Tony Farrell, Greg Burley and the GHOST team for providing the GHOST test data, and Tayyaba Zafar for the AAOmega data. Jon Lawrence, Greg Burley and the anonymous referee made helpful comments on the manuscript.

**REFERENCES**

Barlow, R.J. *Statistics. A Guide to the Use of Statistical Methods in the Physical Sciences* Wiley, p. 78, 1989.

Bevington, P.R. and Robinson, D.K. *Data Reduction and Error Analysis for the Physical Sciences* 2nd ed, McGraw-Hill, 1992.

Eadie, W.T., Dryard, D., James, F.E., Roos, M. and Sadoulet, B. *Statistical Methods in Experimental Physics* North-Holland p. 64, 1971.

Janesick, J.R., Klaasen, K.P. and Elliott, T. Opt. Eng. 26, 972, 1987.

Janesick, J.R. *Photon Transfer. DN $\to \lambda$* SPIE Press, 2007.

Press, W.H., Teukolsky, S.A., Vetterling, W.T. and Flannery, B.P. *Numerical Recipes in C* Cambridge University Press, Sec. 15.2, 1992.

Saunders, W., Bridges, T., Gillingham, P., Haynes, R., Smith, G.A., Whittard, J.D., Churilov, V., Lankshear,

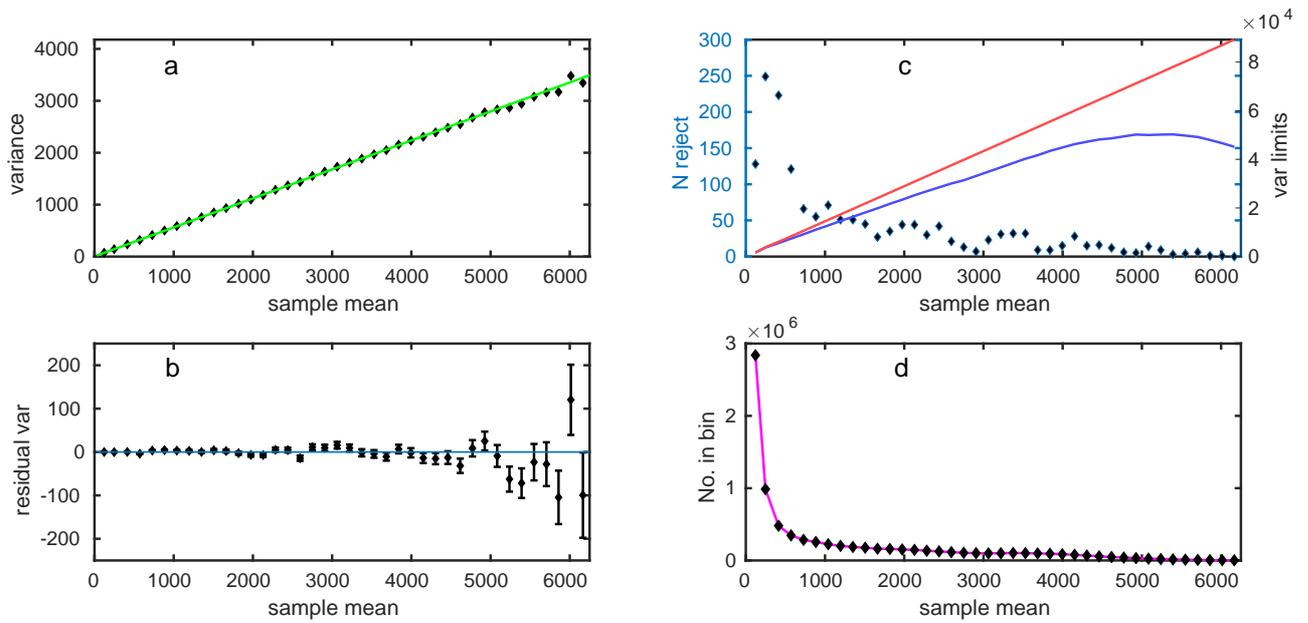

**Figure 6.** Results of analysis of the AAOmega flat-field pair, after correction of the second member of the pair to remove the relative shifts. The panels show the same quantities as in Figure 2. The $s^2$ *vs* $\bar{x}$ line has been fitted over the entire plotted range of $30 - 6250$ ADU.

A., Croom, S., Jones, D. and Boshuizen, C. Proc. SPIE 5492, 389, 2004.

Sheinis, A., Anthony, A., Baker, G., Burley, G., Churilov,V., Edgar, M., Ireland, M., Kondrat, Y., McDermid, R., Pazder, J., Robertson, J., Young, P. and Zhelem, R. Proc. SPIE 990817, 2017.

Stuart, A. and Ord, J.K. *Kendall's Advanced Theory of Statistics* Oxford University Press, Vol. 1. Sec. 16.2, 1987.

Wilson, J.C. et al. PASP 131:055001, 2019.